\documentclass[10pt,showpacs,showkeys,preprintnumbers]{article}
\usepackage{amsmath,amssymb,amsfonts}
\usepackage{units}
\usepackage{bbold,euler}%opcoes de fonte: mathptmx, helvet, eulervm, avant, fourier, newcent, mathpazo,euler,newcent,sansmath
\usepackage{graphicx}% Include figure files
\usepackage{dcolumn}% Align table columns on decimal point
\usepackage{bm,bbm,mathtools,esvect}
\usepackage{slashed}
\usepackage{esvect}
\usepackage{fullpage}
\usepackage[utf8]{inputenc}

\title{\LARGE\bf Quaternionic Dirac free particle}
\author{{\bf Sergio Giardino\footnote{\tt sergio.giardino@ufrgs.br}}\\
\small \it Departamento de Matem\'atica Pura e Aplicada \\
\small \it Universidade Federal do Rio Grande do Sul (UFRGS)\\
\small \it Caixa Postal 15080, 91501-970  Porto Alegre RS \\
\small \it Brazil}
\begin{document}
\date{} % Remove a data
\maketitle

\begin{abstract}
\noindent We solve the quaternionic Dirac equation ($\mathbbm H$DE) in the real Hilbert space, and we ascertain that their free particle solutions set comprises eight elements in the case of a massive particle, and a four elements solution set in the case of a massless particle, a richer situation when compared to the four elements solutions set of the usual complex Dirac equation ($\mathbbm C$DE). These free particle solutions were unknown in the previous solutions of anti-hermitian quaternionic quantum mechanics, and constitute an essential element in order to build a quaternionic quantum field theory ($\mathbbm H$QFT).

\vspace{2mm}

\noindent {\bf keywords:} quantum mechanics;  relativistic wave equations; theory of quantized fields

\vspace{1mm}

\noindent {\bf pacs numbers:} 03.65.-w; 03.65.Pm; 03.70.+k.
\end{abstract}

\maketitle
%\tableofcontents
%%%%%%%%%%%%%%%%%%%%%%%%%%%%%%%%%%%%%%%%%%%%%%%%%%%%%%%%%%%%%%%%%%%%%%%%%%
%%%%%%%%%%%%%%%%%%%%%%%%%%%%%%%%%%%%%%%%%%%%%%%%%%%%%%%%%%%%%%%
\section{\;\bf Introduction\label{I}}
%%%%%%%%%%%%%%%%%%%%%%%%%%%%%%%%%%%%%%%%%%%%%%%%%%%%%%%%%%%%%%%
%%%%%%%%%%%%%%%%%%%%%%%%%%%%%%%%%%%%%%%%%%%%%%%%%%%%%%%%%%%%%%%%%%%%%%%%%%

Quaternionic quantum mechanics ($\mathbbm H$QM) is a mathematically modified quantum theory that replaces the usual complex wave functions with quaternionic wave functions. Recalling that quaternions ($\mathbbm H$) are generalized complex numbers, one can hypothesize that $\mathbbm H$QM generalizes the usual complex quantum mechanics ($\mathbbm C$QM). Detailed mathematical and physical surveys of quaternions may be obtained elsewere \cite{Morais:2014rqc,Rocha:2013qtt,Garling:2011zz,Ward:1997qcn}, and 
here we simply define an arbitrary quaternionic number $\,q\,$ as
\begin{equation}\label{e01}
 q=x_0 + x_1 i + x_2 j + x_3 k, \qquad\mbox{where}\qquad x_0,\,x_1,\,x_2,\,x_3\in\mathbbm{R}.
\end{equation}
The anti-commutative imaginary units $i,\,j\,$ and $\,k\,$ satisfy the general rule
\begin{equation}\label{e02}
e_a e_b =\epsilon_{abc}e_c-\delta_{ab},
\end{equation}
where $e_a$ represents the imaginary unit, $\,\epsilon_{abc}\,$ is the anti-symmetric Levi-Civit\`a tensor, $\,\delta_{ab}\,$ is the Kronecker delta and $\,a,\,b,\,c=\{1,\,2,\,3\}.$
 Quaternions can also be written in symplectic notation, where (\ref{e01}) reads
\begin{equation}\label{e03}
q=z_0+z_1j,\qquad z_0=x_0+x_1i\qquad\textrm{and}\qquad z_1=x_2+x_3i.
\end{equation}
Quaternions are generalized complex numbers, or hyper-complexes, because of their higher number of real degrees of freedom, and one can also expect that a quaternionic generalization of quantum mechanics would accordingly increase their power of describing physical phenomena. However, a consistent $\mathbbm H$QM is not easily obtained, and a book by Stephen Adler \cite{Adler:1995qqm} summarizes the first attempt to get it, although a consistent theory was not accomplished because of the breakdown of the Eherenfest theorem, and the consequent ill-defined classical limit  ({\em cf.} Section 4.4 of \cite{Adler:1995qqm}). The approach of $\mathbbm H$QM contained in Adler's book uses anti-hermitian Hamiltonians and is accordingly called anti-hermitian. Another important characteristic of this approach is  the quaternionic Hilbert space. 
In spite of these difficulties, further efforts are currently being carried out to find alternative formulations to $\mathbbm H$QM, and we quote  \cite{Sapa:2020dqm,Steinberg:2020xvf} by way of example. However, the most consistent impulse to $\mathbbm H$QM in recent times is the development of the real Hilbert space approach \cite{Giardino:2018lem,Giardino:2018rhs},  where the Ehrenfest theorem has been proven, and a diversified collection of results has been obtained \cite{Giardino:2019xwm,Giardino:2016xap,Giardino:2017yke,Giardino:2017pqq,Giardino:2020ztf,Giardino:2020cee,Giardino:2021ofo}. These outcomes within the non-relativistic theory suggested the hypothesis of a relativistic $\mathbbm H$QM, whose first achievement was established by the solution of the quaternionic Klein-Gordon equation ($\mathbbm H$KGE) \cite{Giardino:2021lov}. 

The pathway from the $\mathbbm H$KGE  to  a quaternionic Dirac equation ($\mathbbm H$DE) is the natural course in order to establish a relativistic $\mathbbm H$QM. However, at the first sight, the development of the $\mathbbm H$DE seems to be at a more advanced stage  than the simpler case of the $\mathbbm H$KGE, at least if we consider the number of published articles as the criterion to evaluate the acquired knowledge of the subject. We found only ten articles concerning the $\mathbbm H$KGE ({\it cf.} references of \cite{Giardino:2021lov}) while the $\mathbbm H$DE presents a much higher number of published papers, as we shall see in a moment. However, the situation is only apparent, and the allocation of the $\mathbbm H$DE articles  within several categories shows the accumulated knowledge of relativistic $\mathbbm H$QM as incomplete and even chaotically disposed. In order to clarify the situation, we bring to mind that there are many ways to apply hyper-complex numbers to quantum mechanics, and not all of them generate a novel quantum theory. In this article, we classify quantum theories according to their wave functions and to their Hilbert space. The usual quantum mechanics is defined within the complex Hilbert space and the wave functions are complex as well. On the other hand, the anti-hermitian approach to $\mathbbm H$QM settles over quanternionic wave functions within a quaternionic Hilbert space. In this paper, we adopt quaternionic wave functions within a real Hilbert space. Thus,  let us formally evoke the usual Dirac equation
\begin{equation}\label{e04}
\Big(i\gamma^\mu\partial_\mu-m\Big)\psi=0,
\end{equation}
where $\,\gamma^\mu\,$ are Dirac matrices, $\,m\,$ is the mass constant, and $\,\psi\,$ is a complex wave function. There are several ways to introduce a quaternionic structure into (\ref{e04}).  Recalling that a mathematical classification of hyper-complex Dirac spinors in terms of Clifford algebras is still established \cite{DeAndrade:1999xa,Carrion:2003ve}, we turn to an intuitive way to solve the Dirac equation, and leave the determination of the most suitable mathematical basis to be considered in a future research. 
Remembering the Dirac wave function as a four component spinor, the spinorial wave function can be converted into a scalar quaternionic wave function, taking benefit from the common number of components \cite{Arbab:2010kr,Chanyal:2019zse,Cahay:2019bqp,Venancio:2020qqd}.
Another possibility is to consider the wave function as a complexified quaternion, or biquaternion \cite{Edmonds:1977fp,Berezin:1981tq,Morita:1983kw,Govorkov:1985cg,Lambek:1995ihp,DeLeo:1995ww,DeLeo:1998yi,Morita:2007vc,Rawat:2007vm,Arbab:2017ymu,Bolokhov:2017ndw}. In this case, the four quaternionic components are complex, but the complex unit of the quaternionic components are diverse from the quaternionic imaginary units \cite{Ward:1997qcn}. This kind of approach is also present in David Hestenes' algebraico-geometric approach \cite{Hestenes:1990zbw,Hestenes:2020two}, that interprets geometrically the imaginary unit of (\ref{e04}). We could also use quaternionic formalism to write down the derivative operator $\partial_\mu$ and the Lorentz transformations \cite{Teli:1980mr,Morita:1985rx,DeLeo:2000ik,Lambek:1995ihp,Morita:2007vc,Venancio:2020qqd}. All these approaches preserve the complex Hilbert space. An authentic quaternionic Hilbert space approach concerning the anti-hermititian $\mathbbm H$QM comprises the formal results given in Adler's book \cite{Adler:1995qqm}, and a few applications that can be genuinely considered relativistic $\mathbbm H$QM \cite{Rotelli:1988fc,Adler:1989xf,Davies:1990pm,DeLeo:1995yq,DeLeo:1996ger,DeLeo:2013xfa,DeLeo:2015hza,Giardino:2015iia,Kober:2015bkv,Hassanabadi:2017jiz,Hassanabadi:2017wrt}.
Thus, the complete picture of quaternionic applications to the Dirac equation comprises a meaningful number of quaternionic applications in the complex Dirac equation, and a few examples that comprise the relevant references for the real Hilbert space proposal used in this article. The anti-hermitian attempts are not free from the ill defined classical limit, and the aspiration to obtain quaternionic solutions free of this deficiency motivates this article.

As a final remark, we recollect the linear expansion of  an arbitrary quaternionic function $\,\Psi\,$ in the real Hilbert space \cite{Giardino:2018rhs}
\begin{equation}
\Psi=\sum_{\ell=-\infty}^\infty c_\ell \Lambda_\ell,
\end{equation}
 where $\,c_\ell\,$ are real coefficients and $\,\Lambda_\ell\,$ are unitary quaternionic basis elements. In $\mathbbm C$QM the coefficients and the basis elements of the Fourier expansion are both complex, and in the anti-hermitian $\mathbbm H$QM the coefficients and the basis elements are both quaternionic. The real Hilbert space is endowed with a real valued inner product, and from \cite{Harvey:1990sca} a consistent real inner product between the quaternions  $\Phi$ and $\Psi$ is plainly
\begin{equation}\label{u003}
\langle\Phi,\,\Psi\rangle= \frac{1}{2}\int dx^3\Big[\Phi\Psi^* +\Phi^*\Psi \Big],
\end{equation}
where $ \Phi^*$ and $\Psi^*$ are quaternionic conjugates. This real inner product establishes the quantum expectation value in the real Hilbert space $\mathbbm H$QM, and the consistency demonstrated in the non-relativistic results \cite{Giardino:2016xap,Giardino:2017yke,Giardino:2017pqq,Giardino:2019xwm,Giardino:2020ztf,Giardino:2020cee,Giardino:2021ofo} also encourages	 us to apply it the real Hilbert space $\mathbbm{H}$QM formalism to every quantum system. Let us then consider the relativistic quantum problem we want to study.

%%%%%%%%%%%%%%%%%%%%%%%%%%%%%%%%%%%%%%%%%%%%%%%%%%%%%%%%%%%%%%%%%%%%%%%%%%
%%%%%%%%%%%%%%%%%%%%%%%%%%%%%%%%%%%%%%%%%%%%%%%%%%%%%%%%%%%%%%%
\section{\;\bf Quaternionic Dirac equation\label{D}}
%%%%%%%%%%%%%%%%%%%%%%%%%%%%%%%%%%%%%%%%%%%%%%%%%%%%%%%%%%%%%%%
%%%%%%%%%%%%%%%%%%%%%%%%%%%%%%%%%%%%%%%%%%%%%%%%%%%%%%%%%%%%%%%%%%%%%%%%%%
As defined in \cite{Giardino:2018lem}, the real Hilbert space linear four-momentum operator is, 
\begin{equation}\label{d00}
\widehat{p}_\mu\Psi=\left(\frac{\hslash}{c}\partial_t, -\hslash\bm\nabla\right)\Psi\,i,
\end{equation}
where $\,\Psi\,$ is a quaternionic wave function, and we accentuate the right hand side position of the imaginary unit $\,i\,$ in the operator. Selecting the system of units where  $\,\hslash=c=1,\,$ the generalized linear four-momentum operator \cite{Giardino:2019xwm} is
\begin{equation}\label{d01}
\widehat\Pi^\mu\Psi=\big(\partial^\mu -\mathcal A^\mu\big)\Psi\, i,\qquad\mbox{where}\qquad\mathcal A^\mu=a^\mu i+b^\mu j
\end{equation}
is the gauge potential four-vector, $\,a^\mu\,$ is a real four-vector, $\,b^\mu\,$ is a complex four-vector and consequently $\,\mathcal A^\mu\,$ is a pure imaginary quaternionic four-vector. Therefore, the generalized 
quaternionic Dirac equation will be
\begin{equation}\label{d02}
\left(\widehat{\slashed\Pi} -m\right)\Psi=0,\qquad\mbox{where}\qquad\widehat{\slashed\Pi}=\gamma_\mu \Pi^\mu
\end{equation}
and $\gamma^\mu$ are the $4\times 4$ Dirac matrices, such as
\begin{equation}\label{d03}
\gamma^0=\left[
\begin{array}{cc}
\mathbbm 1 & 0\\
0 & -\mathbbm 1
\end{array}
\right],\qquad
\gamma^\ell=\left[
\begin{array}{cc}
 0 & \sigma^\ell\\
-\sigma^\ell & 0
\end{array}
\right],\qquad\mbox{and}\qquad \ell=\{1,\,2,\,3\}.
\end{equation}
Furthermore, $\mathbbm 1$ represents the identity matrix and $\,\sigma^\ell\,$ represent the Pauli matrices. We stress that the relevant difference between the quaternionic Dirac equation (\ref{d02}) and the usual complex case (\ref{e04}) is the position of the imaginary unit at the right hand side of the wave function defined by the momentum operator (\ref{d01}).
 As a first consistency test of the $\mathbbm H$DE (\ref{d02}), let us obtain the continuity equation for the probability density. We adopt the convention$\,\gamma^0=\beta,\,$  $\alpha^\ell=\gamma^0\gamma^\ell$, and multiply the left hand side of (\ref{d02})  by the adjoint 
\begin{equation}\label{d030}
\overline\Psi=\Psi^\dagger \beta,
\end{equation}
 and the right hand side of (\ref{d02}) by $\,i\,$. The real part of this product gives
\begin{equation}\label{d04}
\partial_\mu\mathcal J^\mu=\overline\Psi\, b_\ell\big(\gamma^\ell-\gamma^{*\ell}\big)j\,\Psi,\qquad\mbox{where}\qquad
\mathcal J^\mu=\overline\Psi \,\gamma^\mu\, \Psi
\end{equation}
is the probability density four-current, and $\gamma^{*\mu}$ is the complex conjugate of $\gamma^\mu$. The conservation of the probability density is violated for $b^\mu\neq 0$. The result is similar to the non-relativistic case ({\it cf.} Section 20 of \cite{Schiff:1968qmq}), where  a complex potential describes a non-conservative probability  current associated to particle generation processes \cite{Giardino:2018lem}, or even to non-elastic scattering phenomena.	Consequently, the continuity equation (\ref{d04}) fulfills our expectations, and the probability density is conserved in the majority of the cases. Let us now propose a quaternionic solution for (\ref{d02}) in terms of the symplectic wave function
\begin{equation}\label{e06}
\Psi=\cos\Theta\,\psi^{(0)}+\sin\Theta \,\psi^{(1)}\,j,
\end{equation}
where $\,\psi^{(\alpha)}\,$ are complex functions, $\,\alpha=\{0,\,1\}\,$ and $\,\Theta\,$ is a real function. 
Using (\ref{d02}) and (\ref{e06}), we have
\begin{eqnarray}
\nonumber &&\cos\Theta\Big(i\slashed\partial+\slashed a -m \Big)\psi^{(0)}-i\sin\Theta\Big(\slashed\partial\Theta\,\psi^{(0)}-\slashed b\,\psi^{*(1)}\Big)+\\
\label{d05}
&&+\Bigg[\sin\Theta\Big(i\slashed\partial+\slashed a+m\Big)\psi^{(1)}+i\cos\Theta\Big(\slashed\partial\Theta\,\psi^{(1)}-\slashed b\,\psi^{*(0)}\Big)
\Bigg]j=0.
\end{eqnarray}
Let us choose
\begin{equation}\label{d06}
\Psi=\cos\Theta\,\exp\Big[ik^{(0)}_\mu x^\mu\Big]u^{(0)}+\sin\Theta\,\exp\Big[ik^{(1)}_\mu x^\mu\Big]u^{(1)}\,j,
\end{equation}
where $u^{(\alpha)}$ are constant spinors, $\,k^{(\alpha)\mu}=\Big(k^{(\alpha)0},\,\bm k^{(\alpha)}\Big)\,$   are constant four vectors, 
\begin{equation}
\Theta=\theta_\mu x^\mu+\Theta_0,
\end{equation}
and $\,\theta^\mu\,$ is a constant four vector with $\Theta_0$ a real constant. If $\,\theta^\mu\neq 0\,$, (\ref{d05}-\ref{d06}) give
\begin{equation}\label{d07}
\Big(\slashed k^{(0)}-\slashed a+m\Big)u^{(0)}=0,\qquad\Big(\slashed k^{(1)}-\slashed a-m\Big)u^{(1)}=0,
\end{equation}
and
\begin{equation}\label{d007}
 \slashed\theta\,\psi^{(\alpha)}-\slashed b\,\psi^{*(\alpha')}=0\qquad\mbox{where}\qquad \alpha\neq \alpha'.
\end{equation}
We observe that equations (\ref{d07}-\ref{d007}) are the farthest point we can reach keeping an absolutely general approach to $\mathbbm H$DE, and further developments will need a choice concerning the quaternionic gauge potential $\mathcal A^\mu$. The above system comprises two complex Dirac solutions combined within the quaternionic structure, and consequently the quaternionic  Dirac particle has a richer and more sophisticated structure than the complex Dirac particle.  Let us consider the simplest	 case.

%%%%%%%%%%%%%%%%%%%%%%%%%%%%%%%%%%%%%%%%%%%%%
\section{Quaternionic Dirac free particle}
%%%%%%%%%%%%%%%%%%%%%%%%%%%%%%%%%%%%%%%%%%%%%

In this case, $\,a^\mu=0\,$ and $\,b^\mu=0,\,$ and nontrivial solutions to (\ref{d07}-\ref{d007}) require singular matrix equations, whose determinants respectively give
\begin{equation}\label{d09}
k_\mu^{(\alpha)}k^{(\alpha)\mu}=m^2,\qquad\mbox{and}\qquad\theta_\mu\theta^\mu=0.
\end{equation}
Using the transformation,
\begin{equation}\label{d09a}
\psi^{(\alpha)}\to e^{\pm i\Theta}\psi^{(\alpha)},
\end{equation}
we have a single $\mathbbm C$DE for each value of $\,\alpha,\,$ and we obtain the constraint
\begin{equation}\label{d10}
k^{(\alpha)\mu}\,\theta_\mu=0.
\end{equation}
Consequently, we may define effective momenta, so that
\begin{equation}\label{d10a}
p^{(\alpha)\mu}=k^{(\alpha)\mu}\pm\theta^\mu,\qquad\qquad p^{(\alpha)}_\mu p^{(\alpha)\mu}=m^2,
\end{equation}
and  a nonzero mass solution in one spatial dimension requires $\,\theta^\mu=0.\,$ The effective momenta permit us to obtain the constraints
\begin{equation}\label{d10b}
k^{(\alpha)0}=\pm\,\bm k^{(\alpha)}\bm\cdot\frac{\bm\theta}{\big|\bm\theta\big|}
\qquad\qquad\mbox{and}\qquad\qquad
\theta^0=\pm\frac{\big|\bm k^{(\alpha)}\bm{\cdot\theta}\big|}{\sqrt{\big|\bm k^{(\alpha)}\big|^2+m^2}}.
\end{equation}
Recalling that $\big|\theta^0\big|=\big|\bm\theta\big|$, we immediately recover (\ref{d09}). However, the first expression of (\ref{d10b}) implies that
 $\big|k^{(\alpha)0}\big|<\big|\bm k^{(\alpha)}\big|,$ what is incompatible to (\ref{d09}) because the mass is real. Therefore, we have the constraints
\begin{equation}
\theta^\mu\neq 0\quad\Rightarrow\quad m=0\qquad\mbox{and}\qquad k^{(\alpha)\mu}=\kappa^{(\alpha)}\theta^\mu
\end{equation}
where $\kappa^{(\alpha)}$ is a real constant. On the other hand, $\Theta=\Theta_0$ admits massive and non-massive solutions. Therefore, there is a single massless solution if $\theta^\mu\neq 0$, and massive and massless solutions if $\theta^\mu=0$.
%%%%%%%%%%%%%%%%%%%%%%%%%%%
%%%%%%%%%%%%%%%%%%%%%%%%%%%
\subsection{ the massive quaternionic Dirac free particle}
%%%%%%%%%%%%%%%%%%%%%%%%%%%
%%%%%%%%%%%%%%%%%%%%%%%%%%%
The quaternionic wave function (\ref{d06}) for $\,\Theta=\Theta_0\,$ will be built in terms of parallel quaternionic wave functions, a method successfully used in	 non-relativistic cases \cite{Giardino:2020cee,Giardino:2021ofo}. The  a parallelism condition is necessary in order to ascertain the orthogonality between quaternionic quantum states, remembering that parallel quaternions
({\it cf.} Section 2.5 of \cite{Ward:1997qcn}) are sucht that
\begin{equation}\label{s01}
\mathfrak{Im}\big[pq^*\big]=0\qquad \mbox{where}\qquad p,\,q\in\mathbbm H.
\end{equation}
The above condition is identical to that used to build Fourier series using complex exponential functions, and the parallel quaternions $p$ and $q$ are such that $\,q=|q|\Lambda,\,$  $\,p=|p|\Lambda\,$ and $\,\Lambda\,$ is an unitary quaternion. In order to establish the notation, we introduce the two component spinors $\,u_S,\,$ where $\,S\in\{\uparrow,\,\downarrow\},\,$ and the up and down elements are such that
\begin{equation}\label{d11}
u_\uparrow=\left(
\begin{array}{c}
1 \\ 0
\end{array}
\right)
\qquad\mbox{and}\qquad
u_\downarrow=\left(
\begin{array}{c}
0 \\ 1
\end{array}
\right).
\end{equation}
From the $\mathbbm C$DE, the up  and down spin states define four possible solutions, where the positive and negative energies can be  associated to both of the spin states. In the quaternionic case, the situation is richer because the  $u^{(\alpha)}$ spinors of (\ref{d06})  are independent, encompassing eight combinations between the spin and the energetic parameters $\,k_0^{(\alpha)},\,$ as we shall see. Defining $\,n_+^{(\alpha)}\,$ and $\,n_-^{(\alpha)}\,$ as normalization constants, the spinors are
\begin{equation}\label{d12}
u_{S+}^{(\alpha)}=n_+^{(\alpha)}\left(
\begin{array}{c}
u_S \\ \\
\frac{\bm k^{(\alpha)}\bm{\cdot\sigma}}{k_0^{(\alpha)}+m} \,u_S
\end{array}
\right),
\qquad\mbox{if}\qquad k_0^{(\alpha)}>0,
\end{equation}
and
\begin{equation}\label{d13}
u^{(\alpha)}_{S-}=n_-^{(\alpha)}\left(
\begin{array}{c}
-\frac{\bm k^{(\alpha)}\bm{\cdot\sigma}}{\big|k_0^{(\alpha)}\big|+m}\, u_S\\ \\
u_S
\end{array}
\right)\qquad\mbox{if}\qquad k_0^{(\alpha)}<0.
\end{equation}
We remark that the flipped signal of the mass terms in (\ref{d07}) imposes flipped signals to the energies associated to the pure complex ($u^{(0)}$) and pure quaternionic ($u^{(1)}$) spinors of the wave function.  Moreover, the spinors are also changed, and it is not possible to have a common constant spinor in each component of the wave function, a fact already observed within the scope of anti-hermitian $\mathbbm H$QM
\cite{DeLeo:2013xfa}.  In accordance with the complex case, the covariant normalization of the spinors is such that 
\begin{equation}\label{d14}
u^{(\alpha)}_{S,\pm}\, u^{\dagger(\alpha)}_{S,\pm}=\mathcal E^{(\alpha)},\qquad\mbox{where}\qquad\mathcal E^{(\alpha)}=\left\{
\begin{array}{l}
\big|k_0^{(\alpha)}\big|,\qquad\mbox{or}\\ \\
\frac{1}{m}\big|k_0^{(\alpha)}\big|.
\end{array}
\right.
\end{equation}
Both of the choices for $\mathcal E^{(\alpha)}$ are valid in the massive case we care considering.
Consequently, the wave function is
\begin{equation}\label{d15}
\Psi^{\pm\mp}_{SS'}=\cos\Theta_0\,\exp\Big[ik^{(0)}_\mu x^\mu\Big]u^{(0)}_{S,\pm}\,+\,\sin\Theta_0\,\exp\Big[ik^{(1)}_\mu x^\mu\Big]u^{(1)}_{S,\mp}\,j,
\end{equation} 
and we stress the flipped signals between the components of the wave function. Accordingly,
\begin{equation}\label{d16}
\Psi^\dagger\,\Psi	=\cos^2\Theta_0 \,\mathcal E^{(0)}+\sin^2\Theta_0 \,\mathcal E^{(1)}.
\end{equation}
We can obtain a quaternionic wave function normalized to the unity following the complex case and
defining the complex components of (\ref{e06}) to satisfy
\begin{equation}\label{d17}
\psi^{(\alpha)}=\int d^4k^{(\alpha)}\, A\left(k^{(\alpha)}\right)\, \delta\Big[\big(k^{(\alpha)}\big)^2-m^2\Big]\exp\Big[ik^{(\alpha)}_\mu x^\mu\Big] u^{(\alpha)}
\end{equation}
Were $\,A\left(k^{(\alpha)}\right)\,$ are the coefficients that will be  chosen to get the normalized wave function. We observe a perfect agreement between the complex and the quaternionic cases, where the quaternionic wave functions generalize the complex wave functions in a consistent way.
Allowing $\,A\left(k^{(\alpha)}\right)\delta\big[\big(k^{(\alpha)}\big)^2-m^2\big]\,$ to be an arbitrary  distribution of momenta, the quaternionic Dirac wave packet can be possibly obtained, and this is an interesting direction for future research.

%%%%%%%%%%%%%%%%%%%%%%%%%%%%%%%%%%%%%%%%%%%%%%%%%%%
\section{Spin of the quaternionic Dirac particle}
%%%%%%%%%%%%%%%%%%%%%%%%%%%%%%%%%%%%%%%%%%%%%%%%%%%

Let us consider the case of arbitrary quaternionic spinors
\begin{equation}\label{s02}
\Psi=\cos\theta \,e^{i\phi}\,U_0+\sin\theta\, e^{i\xi}\, U_1 j\qquad\mbox{and}\qquad \Phi=\cos\theta\, e^{i\phi}\,V_0+\sin\theta\, e^{i\xi}\, V_1 j,
\end{equation}
where $U_\alpha$ and $V_\alpha$ are complex spinors. We have that
\begin{equation}\label{s03}
\Psi\Phi^\dagger=\cos^2\theta \,U_0V_0^\dagger+\sin^2\theta\, U_1 V_1^\dagger \,+\,
\sin\theta\cos\theta\, e^{i\big(\phi+\xi\big)}\Big(U_1 V_0^T-U_0V^T_1\Big) j.
\end{equation}
Therefore, parallel quaternionic spinors impose $\,U_0=V_0\,$ and $\,U_1=V_1,\,$ as expected. 
We can use the concept of parallel quaternionic spinors to obtain the orthonormal set of the solutions of the $\mathbbm H$DE.
The four dimensional complex spinors (\ref{d12}-\ref{d13}) comprise four possibilities, two possible spins for each signal of the energy. The quaternionic case assembles four possibilities for each complex component, but the opposed signal of the mass term in (\ref{d07}) reduces the total quantity of quaternionic Dirac spinors from sixteen to eight possible wave functions, classified according to the spin and the signal of the energy parameter of each complex spinor. Consequently, the complete set amounts to
\begin{equation}\label{s04}
\Psi\in\Big\{\,\Psi_{\uparrow\uparrow}^{+-},\,\Psi_{\uparrow\uparrow}^{-+},\,
\,\Psi_{\downarrow\downarrow}^{+-},\,\Psi_{\downarrow\downarrow}^{-+},\,\Psi_{\uparrow\downarrow}^{+-},\,\Psi_{\uparrow\downarrow}^{-+},
\,\Psi_{\downarrow\uparrow}^{+-},\,\Psi_{\downarrow\uparrow}^{-+}\,\Big\}.
\end{equation}
The above set does not necessarily  contains the simplest basis elements to express the solution, something that may be determined considering the classification of hyper-complex Dirac spinors determined in
\cite{DeAndrade:1999xa,Carrion:2003ve}. However, this question is not simple, and deserves a separate investigation, and therfore we  left it  to be considered in future work. The elements of (\ref{s04}) become an orthogonal set after imposing several conditions.
First of all, the real Hilbert space determines that the inner product is real, and thus, following the inner product defined in the non-relativistic case \cite{Giardino:2018lem,Giardino:2019xwm,Giardino:2018rhs}, we have
\begin{multline}\label{s05}
\left\langle \Psi_{SS'}\left(k^{(0)},\,\pm\,\Big|\,k^{(1)},\,\mp\right),\,\Psi_{RR'}\left(q^{(0)},\,\pm\,\Big|\,q^{(1)},\,\mp\right)\right\rangle=\\
=\cos^2\Theta_0\; U_S^\dagger\left(k^{(0)},\,\pm\right)U_R\left(q^{(0)},\,\pm\right)\,+\,\sin^2\Theta_0 \;U_{S'}^\dagger\left(k^{(1)},\,\mp\right) U_{R'}\left(q^{(1)},\,\mp\right)
\end{multline}
However, the wave functions set (\ref{s04}) is not orthogonal using (\ref{s05}) alone, and further constraints are necessary. For example, $\Psi_{\uparrow\uparrow}$ and $\Psi_{\downarrow\downarrow}$ are mutually orthogonal, as well as 
$\Psi_{\uparrow\downarrow}$ and $\Psi_{\downarrow\uparrow}$, however, $\Psi_{\uparrow\uparrow}$ and $\Psi_{\uparrow\downarrow}$ are not immediately orthogonal. To eliminate this difficulty, both of the complex  components of the quaternionic wave function coequally contribute to the orthogonality condition, a resort that has been successfully applied to obtain orthogonal sets in non-relativistic solutions \cite{Giardino:2020cee,Giardino:2021ofo}. Such kind of condition assures that each real coefficient of the Fourier quaternionic expansion of the wave function in the Hilbert space pertains to a single quaternionic direction, and therefore the parallelism condition (\ref{s01}) is preserved in the expansion.
For the sake that  condition (\ref{s05}) be held when selecting each state, it is necessary that both of the spin states be parallel  to obtain a nonzero inner product between base states, thus it is expected that
\begin{equation}
\big\langle \Psi_{SS'},\,\Psi_{RR'}\big\rangle\,\propto\,\delta_{SR}\,\delta_{S'R'},
\end{equation}
and consequently four independent spin states are generated by the quaternionic parallelism condition. In the particular case when 
$\,\mathcal E^{(0)}=\mathcal E^{(1)}=\mathcal E$, 
\begin{multline}
\int dx^3\left\langle \Psi_{SS'}\left(k^{(0)},\,\pm\,\Big|\,k^{(1)},\,\mp\right),\,\Psi_{RR'}\left(q^{(0)},\,\pm\,\Big|\,q^{(1)},\,\mp\right)\right\rangle=\\
=\mathcal E\,\big(2\pi\big)^6\,\delta_{SR}\,\delta_{S'R'}\, \delta^3\left(k^{(0)}-q^{(0)}\right)\delta^3\left(k^{(1)}-q^{(1)}\right).
\end{multline}
The above expression encompasses all the eight orthogonal states to the quaternionic Dirac wave function for a free particle, and is totally compatible to the complex case.

%%%%%%%%%%%%%%%%%%%%%%%%%%%%%%%%%%%%%%%%%%%%%%%%%%%
\subsection{Orthogonality}
%%%%%%%%%%%%%%%%%%%%%%%%%%%%%%%%%%%%%%%%%%%%%%%%%%%
Using the adjoint wave function (\ref{d030}), we obtain orthogonality conditions
\begin{eqnarray}
\label{h01}
&&\Big\langle \overline\Psi_{SS'}(+-),\,\Psi_{RR'}(+-)\Big\rangle\,=\,\delta_{SR}\,\delta_{S'R'}\cos 2\Theta_0,\\
\label{h02}
&&\Big\langle \overline\Psi_{SS'}(-+),\,\Psi_{RR'}(-+)\Big\rangle\,=\,-\delta_{SR}\,\delta_{S'R'}\cos 2\Theta_0.
\end{eqnarray}
The above result is similar to the complex case, and actually recover it in the cases where $\,\cos 2\Theta_0=\pm 1.\,$ However,  an actually new physical situation arises in the cases where $\,\cos 2\Theta_0=0.\,$
Identically to the complex case, we can express the orthogonality in terms of the helicity $\widehat h$, whose operator  is such as
\begin{equation}
\widehat h=\frac{1}{|\bm k|}\widehat{\bm S}\bm{\cdot k}, \qquad\mbox{and}\qquad \widehat{\bm S}=\frac{1}{2}\tilde{\bm\alpha},
\qquad\mbox{where}\qquad \tilde\alpha_\ell=
\left[ 
\begin{array}{cc}
\sigma_\ell & 0  \\
0           & \sigma_\ell
\end{array}
\right],
\end{equation}
and  $\hat{\bm S}$ is the spin operator. The orthogonality conditions (\ref{h01}-\ref{h02}) are satisfied also in terms of the helicity, where
 each complex component of the quaternionic wave function has their own helicity associated to their spin, and the quaternionic wave function is consequently is such that
\begin{equation}
\Psi_{RR'}\to\Psi_{hh'}.
\end{equation}
Relations (\ref{h01}-\ref{h02}) are similar to the complex case, but the particular case where $\cos 2\Theta_0=0$ is an interesting and novel direction for future research.

\subsection{massless solutions}

The results considered to massive solutions are immediately extended to the massless case in the case of $\Theta=\Theta_0$. The $\mathbbm C$DE has two solutions, let us call them $\psi^{(\mathcal L)}$ and $\psi^{(\mathcal R)}$, and there are four quaternionic solutions according to (\ref{e06}), two for each complex component. As expected and in accordance to the complex case, there are fewer possibilities in the massless case.

%%%%%%%%%%%%%%%%%%%%%%%%%%%%%%%%%%%%%%%%%%%%%%%%%%%
\section{Conclusion}
%%%%%%%%%%%%%%%%%%%%%%%%%%%%%%%%%%%%%%%%%%%%%%%%%%%
The previous solutions of the $\mathbbm H$DE obtained using the anti-hermitian approach are dependent on a quaternionic potential without a clear physical interpretation, and whose comparison to the complex solutions is very difficult. Therefore, the free particle solution of the $\mathbbm H$DE presented in this article is the simplest quaternionic solution ever obtained for this equation.  Using the real Hilbert space approach to $\mathbbm H$QM, the Dirac equation has a consistent probability density current, and the obtained solutions form a complete solutions set, something that were never obtained in the usual anti-hermitian approach to $\mathbbm H$QM.  The quaternionic Dirac free particle presents a clear increase in the degrees of freedom and consequently a higher number of the physical possibilities. Furthermore, the quaternionic helicity has additional possibilities unknown in the complex free particle. Because of these clear advances to the complex case, there are numerous possible directions for future research, including the study of the effect of the gauge potential and the zitterbewegung. However, the most challenging task is clearly the development of the Dirac field in $\mathbbm H$QFT, something that seems to a real possibility after the previous study of quaternionic Klein-Gordon equation \cite{Giardino:2021lov}, and the quaternionic Dirac equation presented in this article.

%%%%%%%%%%%%%%%%%%%%%%
%
%
%  BIBLIOGRAPHY
%
%

\bibliographystyle{unsrt} 
\begin{footnotesize}

\end{footnotesize}

\end{document}